\documentclass[prb,twoside,twocolumn,showpacs,floatfix,letterpaper]{revtex4}

\usepackage{graphicx}
\usepackage{amsmath}
\usepackage{amssymb}

\newcommand{\CeFe}{Ce$_2$Fe$_{17}$}
\newcommand{\RFe}{R$_2$Fe$_{17}$}
\newcommand{\YFe}{Y$_2$Fe$_{17}$}
\newcommand{\ThZn}{Th$_2$Zn$_{17}$}
\newcommand{\ThNi}{Th$_2$Ni$_{17}$}
\newcommand{\TN}{\emph{T$_{\mathrm{N}}$}}
\newcommand{\TT}{\emph{T$_{\mathrm{t}}$}}
\newcommand{\TO}{\emph{T$_{0}$}}

\newcommand{\half}{$\frac{1}{2}$}
\newcommand{\sg}{$R\overline{3}m$}

\newcommand{\ca}{\emph{c}-axis}

\begin{document}

\title{Magnetic phase diagram of \CeFe}
\author{Y. Janssen}
\email[Email: ]{yjanssen@ameslab.gov}
\affiliation{Ames Laboratory US DOE, Ames, IA 50011, USA}

\author{S. Chang}
\affiliation{Ames Laboratory US DOE, Ames, IA 50011, USA}

\author{A. Kreyssig}
\affiliation{Ames Laboratory US DOE, Ames, IA 50011, USA}

\author{A. Kracher}
\affiliation{Ames Laboratory US DOE, Ames, IA 50011, USA}

\author{Y. Mozharivskyj}
\affiliation{Department of Chemistry, McMaster University, Hamilton, Ontario, Canada L8S 4M1}

\author{S. Misra}
\affiliation{Ames Laboratory US DOE and Department of Chemistry, Iowa State University, Ames, IA 50011, USA}

\author{P.~C. Canfield}
\affiliation{Ames Laboratory US DOE and Department of Physics and
Astronomy, Iowa State University, Ames, IA 50011, USA}

\date{\today}
\begin{abstract}
Rare-earth-based permanent-magnet materials rich in iron have relatively low ferromagnetic ordering temperatures. This is believed to be due to the presence of antiferromagnetic exchange interactions, besides the ferromagnetic interactions responsible for the magnetic order. 
The magnetic properties of \CeFe\ are anomalous. Instead of ferromagnetic, it is antiferromagnetic, and instead of one ordering temperature, it shows two, at the N\'{e}el temperature \TN\ $\sim$ 208 K and at \TT\ $\sim$ 124 K. \CeFe, doped by 0.5\% Ta, also shows two ordering temperatures, one to an antiferromagnetic phase, at \TN\ $\sim$ 214 K, and one to a ferromagnetic phase, at \TO\ $\sim$ 75 K. 
In order to clarify this behavior, single-crystalline samples were prepared by solution growth, and characterized by electron microscopy, single crystal x-ray diffraction, temperature-dependent specific heat, and magnetic field and temperature-dependent electrical resistivity and magnetization.
From these measurements, magnetic \emph{H-T} phase diagrams were determined for both Ta-doped \CeFe\ and undoped \CeFe. These phase diagrams can be very well described in terms of a theory that gives magnetic phase diagrams of systems with competing antiferro- and ferromagnetism. 

\end{abstract}
\pacs{75.30.Kz,64.60.Kw,71.20.Lp,61.50.Nw}
\maketitle

\section{Introduction}

Modern permanent magnet materials are intermetallic compounds 
containing rare earths (R) and transition metals, which order 
magnetically at temperatures much above room temperature, 
and combine a large		      																					
ferromagnetic moment with a large easy-axis magnetic                  
anisotropy~\cite{Buschow91}. The																			
large magnetic moment (and the high ordering temperature) is
mainly provided by strongly (and ferromagnetically) coupled itinerant 
transition-metal magnetic moments. The strong magnetic anisotropy 
is mainly provided by the localized 4f-rare-earth magnetic moments,
to which the transition-metal magnetic moments are coupled.			 			

Because of its relative abundance in Earth's crust, Fe is the most	  
favorable (cheapest) magnetic transition metal for use in these				
permanent-magnet materials. Among the rare-earth elements, Ce is
most abundant, and therefore cheapest, potentially making an Fe-rich Ce-Fe compound 
an economical permanent magnet material. However, the 
aforementioned requirements are not met. Instead, the Ce-Fe compound 
richest in Fe, rhombohedral \ThZn-type \CeFe\ shows 
abnormal~\cite{Buschow70} magnetic behavior. However, since \CeFe\ is 
chemically closely related to successfully applied permanent magnet 
materials~\cite{Buschow77,Buschow91}, such as SmCo$_5$ and Nd$_2$Fe$_{14}$B, 
understanding its magnetism may lead to a greater understanding 
of magnetism in rare-earth--transition-metal intermetallic compounds.

The \RFe\ compounds form, dependent on the size of the rare-earth 
ion, in two different but related crystallographic structures,  
rhombohedral \ThZn-type for light (larger) rare earths and hexagonal 
\ThNi-type for heavy (smaller) rare earths~\cite{Buschow66}. Generally, 
they are ferromagnetic (Fe moments parallel to R-4f moments) for light rare 
earths, and ferrimagnetic (Fe moments antiparallel to R-4f moments) 
for heavy rare earths~\cite{Givord74-2}. Exceptions are Lu$_2$Fe$_{17}$ and 
Ce$_2$Fe$_{17}$. These compounds have been reported to show, at 
least in some temperature range, antiferromagnetic behavior 
(see e.\ g.\ Refs.~\onlinecite{Givord74-2, Janssen97, Kozlenko04}), 
which for Lu$_2$Fe$_{17}$ occurs between the magnetic ordering 
temperature (275~K) and a second critical temperature
(140~K), below which it is ferromagnetic. Under a hydrostatic 
pressure of 0.4 GPa, Lu$_2$Fe$_{17}$ remains antiferromagnetic 
at temperatures down to 5~K~\cite{Kamarad05}.

As a representative for the magnetic behavior of the Fe-magnetic 			
subsystem in R-Fe compounds, one often considers the Y-Fe analogs
(La-Fe binary intermetallic compounds do not form), 
and such an analog for \CeFe\ is \YFe. 
\YFe\ can be prepared either with a hexagonal \ThNi-type or with 
a rhombohedral \ThZn-type crystal structure. Hexagonal and 
rhombohedral \YFe\ are very similar in their 
magnetic properties, but hexagonal \YFe\ may be easier to obtain
(see e.\ g.\ Ref.\ \onlinecite{Prokhnenko05}), 
while rhombohedral \YFe\ is easier to model\cite{Coehoorn89}, 
due to its simpler crystal structure. 
Hexagonal \ThNi-type \YFe\ is an easy-plane ferromagnet: it 					
orders magnetically near 310~K~\cite{Givord74-2,Prokhnenko05}, and		
the saturated magnetic moment equals $\sim$~1.95 $\mu_\mathrm{B}$/Fe at 4.2~K.	
In contrast to the magnetic 
behavior of Co-rich Y-Co compounds, Fe-rich Y-Fe compounds have 
magnetic ordering temperatures that \emph{de}crease with 
\emph{in}creasing Fe concentration, which helps put the relatively 
low ordering temperature of \YFe\ in perspective 
(see e.\ g.\ Ref.\ \onlinecite{Buschow77}). 
Electronic-structure calculations indicate that Y-Fe compounds  	   	
can be classified as weak ferromagnets~\cite{Coehoorn89}, with 
incomplete occupation of the majority-spin bands. For \YFe, a rapid 
variation of the density of states near the Fermi 
level~\cite{Coehoorn89} was found. This may explain the calculated
strong (and non-linear) volume dependence of magnetization.						
Experimentally, hexagonal \ThNi-type \YFe\ was found to have a 
negative thermal expansion below its magnetic ordering 
temperature~\cite{Givord74}, which was phenomenologically related to 
strongly distance-dependent positive and negative magnetic exchange
interactions between different Fe-magnetic moments on 
different crystallographic sites, a notion that was confirmed by 
calculations related to electronic structure~\cite{Sabiryanov98}. 
Recently, Prokhnenko et al.\ ~\cite{Prokhnenko05} 
found that hexagonal \YFe\ shows 
antiferromagnetic behavior under sufficiently high pressures. 
They indicate that the magnetism in \YFe\ under pressure may be 
compared to antiferromagnetism in unstable $\gamma$-Fe, which 
is still a challenge for electronic structure 
calculations~\cite{Sandratskii98,Knopfle00,Marsman02,Shallcross06}. 
   
As determined from polycrystalline samples \CeFe\ orders 
antiferromagnetically below \TN\ $\sim$ 208~K
~\cite{Janssen97}. At 125~K, a second magnetic-order
transition is observed, to another \emph{anti}ferromagnetic phase
~\cite{Plumier74,Janssen97,Koyama98,Koyama01,Ishikawa03,Makihara03,Makihara04}.
However, 
there are reports, in which \CeFe\ is found to be \emph{ferro}magnetic 
at temperatures below an antiferro-ferromagnetic transition 
temperature~\cite{Prokhnenko02,Makihara03,Makihara04,Teplykh04}, 
which is sample dependent and ranges between $\sim$ 20 and 
$\sim$ 120~K~\cite{Teplykh04}. It has also been 
reported~\cite{Middleton95}, that \CeFe\ doped with $\sim$1.4\% 
Si is ferromagnetic below $\sim$ 120~K, and antiferromagnetic 
between $\sim$ 120~K and $\sim210$~K. Based on their M\"{o}ssbauer 
spectroscopy experiments on different samples, nominally  undoped 
or doped with Si or Al, from different laboratories, Hautot 
et al.\ ~\cite{Hautot00} found that the samples which are 
ferromagnetic at low temperatures may be doped by 
small amounts of Si or by Al. Such doping may be 
caused by samples reacting with Al$_2$O$_3$ crucibles~\cite{Buschow77,Janssen05} 
or silica ampoules. 

Based on the lattice constants, the unit cell of \CeFe\ is smaller than expected from La-contraction, 
Buschow and van Wieringen~\cite{Buschow70} deduced that Ce in \CeFe\ is tetravalent,
and therefore carries no local 4f moment. Later X-ray absorption spectroscopy, 
by Isnard et al.~\cite{Isnard94,Isnard96,Isnard97} and Vandormael et 
al.~\cite{Vandormael97} indicated that Ce in \CeFe\ is in a mixed-valent state.
Furthermore in substitutionally~\cite{Isnard97,Vandormael97} and 
interstitially~\cite{Isnard94,Isnard96} modified \CeFe,  the spectroscopic 
Ce valence decreases with increasing Ce site volume.

Powder-neutron diffraction experiments on apparently undoped \CeFe~\cite{Plumier74,Makihara04} 
indicated that below \TN\ $\sim$ 208~K, \CeFe\ orders in a complex AF structure, which is modified below
\TT\ $\sim$ 125~K. In both cases the magnetic moments were found to order perpendicular to the (hexagonal notation) \ca. 
These powder-neutron results will be discussed elsewhere in this volume~\cite{Kreyssig06}.

Magnetization experiments on polycrystalline samples hinted at a rich magnetic 
(\emph{H-T}) phase diagram \cite{Janssen97,Koyama01}, that is strongly pressure 
dependent~\cite{Ishikawa03}, and involves strong magnetovolume effects~\cite{Makihara03}.
Recently, Makihara et al.\ ~\cite{Makihara04} 
succeeded in preparing single crystals of apparently undoped \CeFe, that displayed 
anisotropic magnetization. 
However, a magnetic phase diagram of single-crystalline \CeFe\ has not been determined. 

Here we report on solution-grown single crystals of 
\CeFe\ and of Ta-doped \CeFe, and on the characterization
of the crystals by magnetization, electrical transport, specific heat, 
x-ray diffraction, and electron-probe micro analysis. 
Below, we will first study the composition and the 
crystal structure of Ta-doped crystals and compare them to undoped crystals. 
Then we present the results of extensive thermodynamic and transport measurements 
to determine the anisotropic \emph{H-T} phase diagrams for undoped \CeFe. After this, we 
present an \emph{H-T} phase diagram obtained on a Ta-doped crystal of \CeFe. Finally, 
the \emph{H-T} phase diagrams are discussed in terms of a theory of competing ferro- and antiferromagnetic interactions.

\section{Experimental}
\label{exp}

Traditionally, R-Fe compounds are grown by self-flux out of Ta crucibles~\cite{Canfield92,Canfield01}.
This technique has been used to grow Nd$_2$Fe$_{17}$ as well as Nd$_2$Fe$_{14}$B single 
crystals~\cite{Koyama96,Lewis98}.
Unfortunately, as will be discussed below, \CeFe\ allows for a slight Ta uptake~\cite{Makihara04} 
on one of the Fe-crystallographic sites, which profoundly alters the magnetic behavior of the sample. 
In order to fully control the amount of Ta in the sample, we have 
used MgO crucibles (Ozark Technical Ceramics, Inc.), as described in Ref.~\onlinecite{Makihara04}, to 
grow single crystals of \CeFe\ as well as of intentionally Ta-doped \CeFe. 

Starting alloys were made by arc melting pieces of Fe(99.98\%, SCM Metal Products Inc., Cleveland, OH), 
Ce (99.9\% elemental, 99.98\% metals basis, Materials Preparation Center, Ames Laboratory~\cite{MPC}) 
and Ta (99.9\%) in a standard arc furnace under a partial pressure of $\sim$~0.8 bar of Ti-gettered 
ultra-high-purity Ar. Prior to the crystal growth experiments, the composition and temperature range 
for crystal growth were optimized with the aid of differential thermal 
analysis~\cite{Janssen05} (DTA). 

The button of the starting alloy was placed in the growth crucible, a 2 ml MgO crucible. 
This growth crucible was capped with a strainer, a piece of Ta foil with several holes in it, 
while care was taken that the foil did not come in contact with the alloy, to prevent possible 
dissolution of the Ta. On top of the capped MgO crucible, a catch crucible, an inverted 2 ml
Al$_2$O$_3$ crucible, was placed. This assembly was sealed in an amorphous silica ampoule 
under $\sim$ 0.3 bar of ultra-high-purity Argon. 

To obtain undoped \CeFe, an alloy with composition 
Ce$_{0.475}$Fe$_{0.525}$ was heated to $1200^{\circ}{\rm C}$, then allowed to equilibrate 
for $\sim$ 2h, quickly cooled to 1000$^{\circ}{\rm C}$, and then cooled slowly to
$940^{\circ}{\rm C}$, where the flux was decanted. The liquid alloy had visibly reacted with 
the crucible: the crucible had become brown, and on the bottom, apparently up to the level 
of the liquid, a brownish crust had formed. On this crust, which contained Ce oxide, 
we also found crystals of \CeFe, which we removed mechanically. As we will see below, no Mg was found 
in these crystals, as may be expected from the limited solid solubility 
of Mg in Fe~\cite{Okamoto00}.

Crystals typically grow blocky-rhombohedral or tabularly, with clean and clear facets, 
see Fig.~\ref{crystalfig}. For the tabular crystals, the facets with largest surface area were 
found perpendicular to the hexagonal \ca, as confirmed by x-ray diffraction. 
(From here, we will use hexagonal notation for the rhombohedral structure.) 
It appeared that the probability to obtain tabular (or even plate-like) 
crystals is larger when the alloy was cooled more quickly, and conversely, 
the probability to obtain blocky (more 3-D)  crystals is larger
when the alloy was cooled more slowly. Therefore, cooling rates varying 
between $0.3-2^{\circ}{\rm C} / {\rm h}$ were used. (Note that the 
DTA-optimization~\cite{Janssen05} is useful here: 
slow cooling was performed over only 60${^\circ}{\rm C}$.)
Among the blocky crystals we could not find crystals longer, in the \emph{c}-direction, 
than $\sim 1.5$ mm. Moreover, crystals reasonably long in the \emph{c}-direction 
appeared to have inclusions that became visible when the crystals were cut. 

Quantitative electron-probe microanalysis (EPMA) was performed
with a JEOL JXA-8200 Superprobe with a 20 kV acceleration
potential and a 20 nA beam current. Crystals of undoped \CeFe\ and
of Ta-doped \CeFe\ were polished and mounted side by side. As a standard for
Ce, a single crystal of the air-insensitive binary line compound
CeRu$_2$ was used. For Mg, a sample of Mg$_2$SiO$_4$ was used as a
standard, and for Fe and Ta, pieces of those elements were used as
standards.

Single-crystal x-ray diffraction measurements were performed on
the undoped and Ta-doped \CeFe. Crystals with diameter $\sim$
10 $\mathrm{\mu}$m$^3$ were obtained by crushing crystals from the
respective batches. Room-temperature x-ray diffraction data were
collected on a STOE IPDSII image plate diffractometer with Mo
K$\alpha$ radiation, and were recorded by taking 1$^\circ$ scans
in $\theta$ in the full reciprocal sphere. 2$\theta$ ranged from
6$^\circ$ to 63$^\circ$. Numerical absorption corrections for both
crystals were based on crystal face indexing, followed by a
crystal-shape optimization. Structure solution and refinement were
done using the SHELXTL program.

Magnetization measurements at temperatures between 2~K and 300~K in applied fields 
of up to 5.5 or 7~T were performed in Quantum Design MPMS-5 and MPMS-7 magnetometers, 
respectively, on plate-like crystals as well as on blocky crystals. Zero-field 
ac-susceptibility was measured in a Quantum Design MPMS-5 magnetometer, 
excitation frequency 1000 Hz, amplitude 0.2 mT, on a tabular crystal of $\sim$ 97 mg. 
Specific heat was determined in a Quantum Design PPMS system on the same crystal as 
the one used for ac-susceptibility. For electrical transport measurements,
we used a standard four-probe technique.  
Bars were cut with a wire saw, 
both for the current \emph{j} flowing parallel to the \ca\ and for the current 
\emph{j} flowing perpendicular to the \ca, and electrical contact was made 
with Epo-tek H20E silver epoxy, with typical contact resistances of about 4 $\Omega$.
To obtain a bar for \emph{j}$\bot$ \ca\ was relatively easy: a plate-like crystal
$\sim$ 0.15 mm thick, and several mm diameter was used to cut a bar of 3 mm 
length and 1 mm width. To obtain a good sample for \emph{j}// \ca\ was much more
difficult, because of the limited available length. 
The resulting sample was 0.4 mm thick, 0.7 mm wide and 1.1 mm long.

\section{Chemical and structural characterization}
\label{chemstruct}

A total of four single-crystal samples were examined by EPMA. For one of these, the starting 
alloy nominally consisted only of Ce and Fe, and for the other three,
the starting alloy was doped with Ta (by 0.05\%, 0.5\% and 1\%).
The nominally undoped sample was very carefully checked for both Mg and for Ta  by performing 
slow wavelength-dispersive spectroscopy scans over the Mg-K and Ta-L lines, 
respectively, on the standards as well as on the sample. 
The spectra from the nominally undoped sample showed no discernible Mg or Ta peaks. 
Spectra from Ta-doped samples did produce Ta peaks, and Ta-peak heights obtained by scanning were consistent 
with Ta-peak heights obtained by a three-point analysis. 
The analysis indicated that the undoped sample contained less than $\sim$ 300 ppm Mg 
and less than $\sim$ 300 ppm Ta.
Quantitatively, the composition of the undoped sample was found equal to Ce$_2$Fe$_{16.9(1)}$.
The nominally 0.05\%, 0.5\%, and 1\% Ta-doped samples were found to contain 0.42(5)\%, 0.37(5)\%, and 0.57(4)\% Ta, respectively,
which indicates that there is a solid-solubility maximum for Ta in \CeFe.
Quantitatively, the compositions were found equal to Ce$_2$Fe$_{17.1(2)}$Ta$_{0.08(1)}$, 
Ce$_2$Fe$_{17.0(1)}$Ta$_{0.07(1)}$, and  Ce$_2$Fe$_{17.00(9)}$Ta$_{0.108(7)}$,
for nominally 0.05\%, 0.5\% and 1\% Ta, respectively. Note that these  
compositions do not differ significantly, i.e.\ by more than two standard deviations, 
from (2-17) stoichiometry. EPMA also indicated that samples may have parasitic inclusions 
of CeFe$_2$, which was also noted in Ref.~\onlinecite{Makihara03}.

\begin{figure} 
\begin{center}
\includegraphics[bb=0 0 4.25in 5.5in,viewport=1.75in 1in 4.25in 5.5in, scale=0.85]{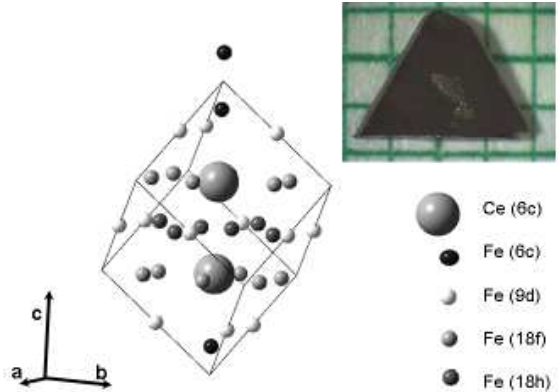}
\caption{Schematic drawing~\cite{Ozawa04} of the rhombohedral unit
cell of \CeFe, with hexagonal axes. In Ta-doped \CeFe\ the Fe (\emph{6c}) site is partially
occupied by Ta. The inset shows a photograph of a plate-like crystal on a mm-grid background. The \emph{c}-direction is perpendicular to this grid, and the sides coincide with main planar directions.}\label{crystalfig}
\end{center}
\end{figure}

\label{diffraction}
\begin{table}
\begin{center}
\begin{tabular}{cccccc}
\hline\hline Atom(site) & \textit {x/a} & \textit {y/b} & \textit
{z/c}  & occ.& \textsl{U}$_{eq}$  \\
\hline
Ce($6c$) & 0 & 0 & 0.34351(3) & 1 & 0.0077(2) \\
Fe($6c$) & 0 & 0 & 0.09682(8) & 1 & 0.0074(3) \\
Fe($9d$) & \half & 0 & \half & 1 & 0.0067(2) \\
Fe($18f$) & 0.29050(8)& 0 & 0 & 1 & 0.0086(2) \\
Fe($18h$) & 0.50150(4) & 0.49850(4)& 0.15502(5) & 1 & 0.0078(2) \\
\hline\hline
\end{tabular}
\caption{Atomic coordinates and equivalent isotropic displacement
parameters (\AA$^2$) for \CeFe. U$_{eq}$ is defined as one third
of the trace of the orthogonal U$^{ij}$ tensor. Space group \sg, \emph{a}
= 8.4890(12) \AA , \emph{c} = 12.410(4) \AA, \emph{R} = 0.0149, \emph{R}$_\mathrm{w}$ = 0.0361.
\label{UndopedXRD}}
\end{center}
\end{table}

\begin{table}
\begin{center}
\begin{tabular}{cccccc}
\hline \hline Atom(site) & \textit {x/a} & \textit {y/b} & \textit
{z/c}
    & occ.& \textsl{U}$_{eq}$  \\
\hline
Ce($6c$) & 0 & 0 & 0.3436(1) & 1 & 0.0068(3) \\
Fe($6c$) & 0 & 0 & 0.0965(2) & 0.954(9) & 0.0074(11) \\
Ta($6c$) & 0 & 0 & 0.0965(2) & 0.046(9) & 0.0074(11) \\
Fe($9d$) & \half & 0 & \half & 1 & 0.0060(5) \\
Fe($18f$) & 0.2915(2)& 0 & 0 & 1 & 0.0071(4) \\
Fe($18h$) & 0.5012(1) & 0.4988(1)& 0.1549(1) & 1 & 0.0080(4) \\
\hline\hline

\end{tabular}
\caption{Atomic coordinates and equivalent isotropic displacement
parameters (\AA$^2$) for Ta-doped \CeFe. U$_{eq}$ is defined as
one third of the trace of the orthogonal U$^{ij}$ tensor. Space
group \sg, \emph{a} = 8.4972(12) \AA , \emph{c} = 12.447(3) \AA, \emph{R} = 0.0496,
\emph{R}$_w$ = 0.0535. \label{TadopedXRD}}
\end{center}
\end{table}

An undoped crystal and the EPMA-crystal, grown out of the nominally 0.05\%
Ta-doped alloy were used for single-crystal X-ray diffraction experiments.
The crystallographic and structural data are summarized in 
Tables~\ref{UndopedXRD} and ~\ref{TadopedXRD}, for undoped and 
Ta-doped \CeFe, respectively. 
We refined the x-ray diffraction-data 
for the undoped sample assuming only Ce and Fe form the structure. 
The diffraction data indicates the crystal has the rhombohedral 
Th$_2$Zn$_{17}$-type crystal structure with space group \sg, as shown in 
Fig.~\ref{crystalfig}, with a full occupancy of all
available sites. The high quality of the refinement (the
refinement factor $R$ = 0.0149), indicates that this crystal has
no obverse-reverse twins (reticular merohedry), as found for the
related compound Pr$_2$Fe$_{17}$, see Ref.~\onlinecite{Calestani03}. We found for
the lattice parameters of the undoped \CeFe\ crystal \emph{a} =
8.4890(12) \AA, and \emph{c} = 12.410(4) \AA. These unit-cell parameters
agree very well with those found for apparently undoped
\CeFe, see Ref.~\onlinecite{Makihara03}. The atomic position parameters, see
Table~\ref{UndopedXRD}, are in good agreement with published
values~\cite{Isnard90}.

Although it was clear from the EPMA experiments that there was Ta in
the doped sample, we initially refined the diffraction data from this
crystal assuming that only Ce and Fe form the structure. Although in this case 
the \emph{R}-factor was only slightly higher than the \emph{R}-factor for the final refinement
shown in Table~\ref{TadopedXRD}, 0.0504 instead of 0.0496, the refined isotropic 
displacement parameter for the Fe-\emph{6c} site became very small, 0.0021(6) instead of 0.0074(11). 
This value is much lower than the displacement parameters found for the 
other Fe atoms in the structure, whereas for pure \CeFe, 
Table~\ref{UndopedXRD}, values were about equal for the different Fe sites.
Since small thermal displacements have the same effect
on diffracted beam intensities as larger electron densities, it appears that 
the electron density on the Fe-\emph{6c} site is too low in the model, without Ta. 
The introduction of an Fe/Ta mixture on the Fe-\emph{6c} site brings the temperature 
factor to the level of the other Fe atoms. 
Further refinements, allowing for mixtures on the other crystallographic 
sites, made clear that Ta has a strong preference to substitute for Fe on 
this Fe-\emph{6c} site. Therefore, the structure was refined with Ta only on that site. 
The final refinement factor $R$ = 0.0496, indicates a good fit. 
Residual electron density not described by the current structural model 
does not lead to Ta occupying possible interstitial sites by more than about 0.5\%.
The refined amount of Ta leads to an empirical x-ray formula of
Ce$_2$Fe$_{16.91(2)}$Ta$_{0.09(2)}$ , which is, within experimental error, 
the same as the composition found by the EPMA experiment for this crystal, 
Ce$_2$Fe$_{17.1(2)}$Ta$_{0.08(1)}$.

The lattice parameters and unit-cell parameters for the Ta-doped \CeFe\
crystal: space group \sg, \emph{a} = 8.4972(12) \AA , \emph{c} = 12.447(3) \AA,
are slightly larger than the lattice parameters for the undoped
crystal, which is consistent with reported~\cite{Makihara03} lattice parameters
for Ta-crucible grown \CeFe.

The result that Ta substitutes for Fe on the Fe-\emph{6c} site, is consistent with expectations,
and the results that Ta has a solid-solubility maximum in \CeFe, is both consistent with 
expectations and plausible.
All transition metals with a radius larger than Fe, which includes Ta, 
Ref.~\onlinecite{Shannon76}, prefer to substitute for Fe on the Fe-\emph{6c} 
site~\cite{Girt98} in Nd$_2$Fe$_{17}$ 
with the same crystal structure as \CeFe. 
The solid solubility of Ta in isostructural Sm$_{2}$Fe$_{17}$ 
has~\cite{Saje95} a maximum of 2.3 at\% of Ta in Sm$_2$Fe$_{17-x}$Ta$_x$.
In solution growth~\cite{Canfield92,Canfield01,Janssen05}, 
the composition of the product (crystals) is different than
the composition of the initial melt (crystals plus solvent).  
Here, the mass ratio (yield) of crystals-to-flux was typically about 10\%,  
which makes it theoretically possible to obtain crystals with as high as 
0.5 at\% Ta for an initial melt with 0.05 at\% Ta. 

The three Ta-doped crystals under investigation are all doped by about 0.5 \% Ta, 
which indicates that there exists, besides a limited solid solubility of Ta in \CeFe,
a mechanism that favors Ta to dope the crystals (\CeFe) rather than remain in
the solvent ($\sim$ Ce$_{0.5}$Fe$_{0.5}$). This mechanism might be related to the 
limited solubility of Ta in liquid Ce ($\sim$ 0.005 \% at $\sim$ $1000^{\circ}{\rm C}$), 
see Ref.~\onlinecite{Dennison66}.

\section{Pure \CeFe- Zero-field results}
\label{Zero-fielddata}

\begin{figure}[!ht]
\begin{center}
\includegraphics[bb=0 0 3.5in 7in,scale=0.9]{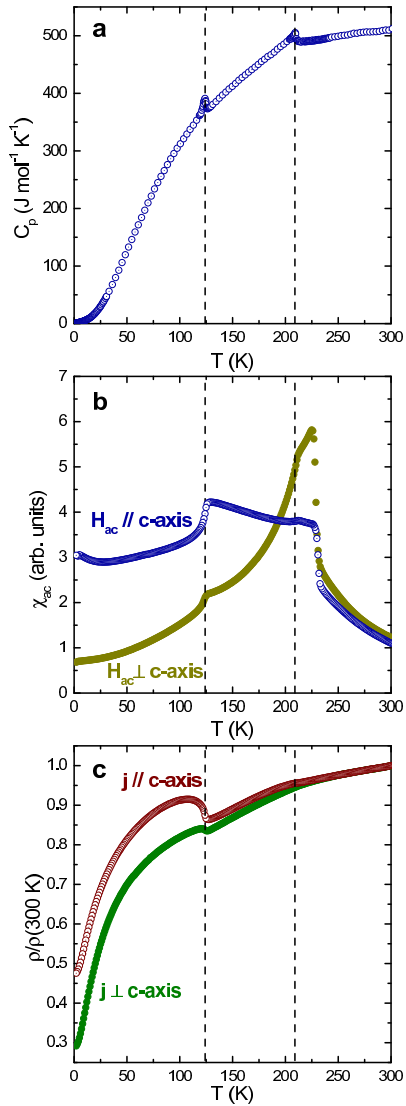}
\caption{\textbf{a} Zero-field specific heat C$_p$ as a function of
temperature. The $\lambda$-type peaks, at \TT\ = 124(1) and \TN\ = 209(2)~K, are
due phase transitions, as described in the text. 
\textbf{b} ac-susceptibility for the ac-field applied both $\bot$ \ca\ and // \ca.  
The sharp rise, in both curves, near 230~K is due to parasitic CeFe$_2$. 
The dashed lines indicate the temperatures where
peaks were found in the specific heat. 
\textbf{c} Normalized temperature-dependent zero-field resistivity
$\rho$(T) for the current flowing parallel to the \ca\ (open
circles), and for the  current flowing perpendicular to the \ca\
(closed circles). The dashed lines indicate the temperatures where
peaks were found in the specific heat. }\label{zerofielddata}
\end{center}
\end{figure}

Temperature-dependent specific heat of \CeFe\ is presented in 
Fig.~\ref{zerofielddata}\textbf{a}. Two $\lambda$-type peaks, at 124~K and at 209~K,
are clearly seen. These peaks are consistent with magnetic ordering
transitions found earlier in polycrystalline samples~\cite{Janssen97}. 

Results of anisotropic zero-field ac-susceptibility measurements 
for both \emph{H}$_\mathrm{ac}$ $\bot$ \ca\ and \emph{H}$_\mathrm{ac}$ // \ca,
are presented in Fig.~\ref{zerofielddata}\textbf{b}.  For both ac-field
directions, three features characterize 
the data: a sharp rise near 225~K, and two weaker anomalies, 
near 208~K and 125~K, respectively. The temperature of the sharp rise, $\sim$ 225~K, 
is consistent with the ferromagnetic ordering temperature of $\sim 230$~K of 
CeFe$_2$~\cite{Buschow70}. (Inclusions of CeFe$_2$ were observed
in samples used for EPMA, see above.) 
Moreover, the sample used for the ac-susceptibility presented here
was the same as the one used for the specific heat experiments described above. 
In that experiment, no anomaly was observed near 225~K.
From a magnetization isotherm measured at 5~K (not shown) we estimate 
the amount of CeFe$_2$ in this sample as less than 0.5\%. 
As for the other two anomalies, they occur at temperatures near 
the temperatures of the peaks in specific heat, and are consistent 
with the previously reported ordering temperatures for \CeFe. 

Results of zero-field resistivity measurements 
are presented in Fig.~\ref{zerofielddata}\textbf{c}. Because of the small 
dimensions of the samples, the estimated error in the determination 
of the resistivity was relatively large, about 20 \%. At room 
temperature, above the magnetic phase transition temperatures, 
overlapping values were found, for \emph{j} // \ca\ and for 
\emph{j} $\bot$ \ca, 150 (30) $\mu \Omega$ cm and 190 (38)$\mu 
\Omega$ cm, respectively. Because of this overlap, the existence 
of anisotropy in the resistivity at room temperature is not clear. 
To further study possible anisotropic temperature dependence,
the resistivity results were normalized to the 300~K values. 

In polycrystalline samples, resistivity anomalies associated with the magnetic 
ordering near 209~K were not observed~\cite{Janssen97,Vandormael97,Koyama98}.
Here, above 220~K, the behavior of the resistivity appears
indistinguishable for both employed current directions. Below that
temperature, thus near 209~K, the temperature of the
highest-temperature transition found in specific heat, the
normalized zero-field resistivity for \emph{j} $\bot$ \ca\
is lower than for \emph{j} // \ca. Thus near the magnetic 
ordering temperature, the resistivity behavior becomes anisotropic.  
Both curves appear approximately linear with temperature between 
$\sim$ 130~K and $\sim$ 190~K.

At a still lower temperature of $\sim$ 124~K, consistent with the
$\lambda$-type peak in specific heat, an anomalous increase of the
resistivity occurs, for both current directions, similar to the 
resistivity behavior below the ordering temperatures of heavy rare earths
like Er or Tm (see e.\ g.\ Refs.~\onlinecite{Ellerby00} 
and~\onlinecite{Ellerby98}, respectively). Such an increase, 
though much less pronounced, was observed earlier in 
polycrystalline \CeFe\ \cite{Janssen97,Koyama98}. The increase is much 
larger for \emph{j} // \ca\ than for \emph{j} $\bot$ \ca. This is consistent 
with the formation of anisotropic superzone gaps, gapping parts of
the Fermi surface, particularly $\bot$ \ca. 

We estimate the gapped fraction of the Fermi surface by considering the conductivity~\cite{Ziman63}

\begin{equation} \label{Kubo}
\sigma=\frac{e^2}{12 \pi^3 \hbar} \int \Lambda d S_{\mathrm F}
\end{equation}

\noindent with $\Lambda$ the electron mean free path, and the integral over 
the Fermi surface area $S_{\mathrm F}$. Assuming that $\Lambda$ does not 
change anomalously due to the transition, the gapped fraction of the Fermi surface is
proportional to the relative reduction $\delta$ in conductivity $\sigma$. 
If we further assume that all parts of the Fermi surface contribute equally to conductivity,
the gapped fraction is equal to the relative reduction in conductivity.

Then the assumed resistivity without superzone gaps, $\rho_\mathrm{without}$, just below 125~K , 
follows the apparently linear behavior observed above 125~K.  
This resistivity without superzone gaps below 125~K is obtained by linear 
extrapolation of the resistivity between $\sim$ 130 and 180~K. An estimate for $\delta$ 
is obtained from Eq.~\ref{Kubo}

\begin{equation}
\delta = 1- \frac{\rho_\mathrm{without}}{\rho_\mathrm{with}}
\end{equation}

The reduced resistivity for \emph{j} // \ca\ shows a maximum near 107~K, 
and we assume for our estimate that the gaps have fully developed there. 
The reduced resistivity at 107~K reads $\rho_\mathrm{with,red}$=0.92, and the 
extrapolated reduced resistivity reads $\rho_\mathrm{without,red}$=0.83. 
Therefore, we estimate that the conductivity for \emph{j} // \ca\ is reduced by 
$\delta \sim$ 10 \% due to the superzone gaps. Obtained in a similar manner, we  
estimate that the conductivity for \emph{j} $\bot$ \ca\ is reduced by $\sim$ 3\%.
This anisotropic behavior is consistent with expectations due to the 
reported propagation vectors for the magnetic structure, parallel to the 
\ca~\cite{Givord74,Fukuda99,Makihara04}.

Furthermore, the zero-field resistivity for both current
directions shows a rather small residual resistance ratio (RRR),
about 2 for \emph{j} // \ca\ and about 3 for \emph{j} $\bot$ \ca.
Such values may indicate poor sample quality, however, because
in an applied field RRR is much larger for both current
directions, as will be shown below, we think
that in \CeFe\ a strong magnetic scattering is the major cause for such 
low RRR values.

\begin{figure}[!ht]
\begin{center}
\includegraphics[bb=0 0 3.5in 3.5in,scale=0.9]{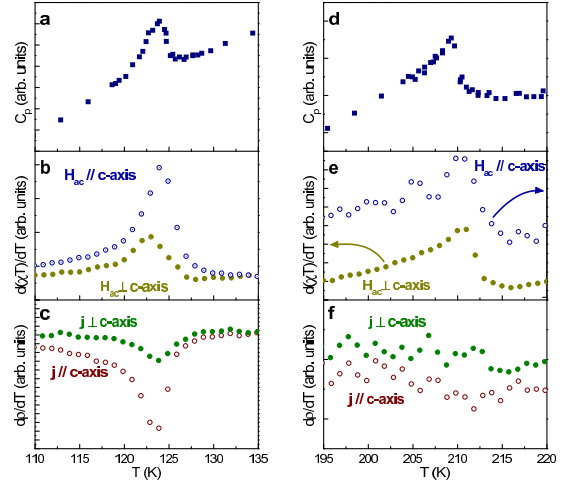}
\caption{Left: \textbf{a} Specific heat,
\textbf{b} $\partial(\chi T)/\partial T$ for H$_{ac} \bot$ \ca\ (closed circles) and for H$_{ac} //$ \ca\ (open circles) , and  
\textbf{c} $\partial \rho / \partial T$ for the current \emph{j} // \ca\ (open
circles), and for \emph{j} $\bot$ \ca\ (closed circles), between 110 and 135~K.
Right: 
\textbf{d} Specific heat,
\textbf{e} $\partial(\chi T)/\partial T$ for H$_{ac} \bot$ \ca\ (closed circles) and for H$_{ac} //$ \ca\ (open circles) , and  
\textbf{f} $\partial \rho / \partial T$ for the current \emph{j} // \ca\ (open
circles), and for \emph{j} $\bot$ \ca\ (closed circles),
between 195 and 220~K.}\label{zfderiv1}
\end{center}
\end{figure}

As already indicated above, in the measurements of specific heat, 
ac-susceptibility and resistivity, the two ordering temperatures 
are observed. Anomalies in $\frac{\partial\rho}{\partial T}$ coinciding with 
specific heat anomalies are frequently observed in magnetic materials 
(see e.g.\ Refs.~\onlinecite{Ribeiro03,Morosan04,Angst05}).
For an accurate phase-diagram determination to be described in the next sections, 
criteria for determining transition temperatures are needed.
In Fig.~\ref{zfderiv1}, we present temperature-dependent specific 
heat (\textbf{a}) between 110 and 135~K, compared to $\frac{\partial(\chi T)}
{\partial T}$ (\textbf{b}) and $\frac{\partial\rho}{\partial T}$ for both 
current directions (\textbf{c}).
There is a good agreement between the specific 
heat and $\frac{\partial(\chi T)}{\partial T}$ (\textbf{b}), even though, 
strictly, the theories in Refs.~\onlinecite{Fisher62} 
and~\onlinecite{Wolf64}, are only good for paramagnetic to 
antiferromagnetic transitions. Clear anomalies, with (besides the fact that they 
point downward) similar shapes as the peak in specific heat, are also observed 
in $\frac{\partial\rho}{\partial T}$ (\textbf{c}) for both current-flow
directions. 

In Fig.~\ref{zfderiv1}, we also present temperature-dependent specific 
heat (\textbf{d}) between 195 and 230~K, compared to $\frac{\partial(\chi T)}
{\partial T}$ for both ac-field directions
(\textbf{e}) and $\frac{\partial\rho}{\partial T}$ for both current directions 
(\textbf{f}). 
Aside from the sharp downturn above $\sim$ 224~K, (see Fig.~\ref{zerofielddata}\textbf{b}), 
which we ascribed to an impurity of CeFe$_2$, there is a good agreement 
between the specific heat (\textbf{d}) and $\frac{\partial(\chi T)}{\partial T}$ 
for both \emph{H} $\bot$ \ca, and reasonable agreement for \emph{H} // \ca (\textbf{e}). 
Such an agreement is expected for simple antiferromagnets as well as for 
less-simple antiferromagnets, see Refs.~\onlinecite{Fisher62} 
and~\onlinecite{Wolf64}, respectively. As expected from the absence of
a clear anomaly in the individual resistivity curves for 
both \emph{j} // \ca\ and \emph{j} $\bot$ \ca, see 
Fig.~\ref{zerofielddata}\textbf{c}, a clear anomaly is not observed in 
$\frac{\partial\rho}{\partial T}$ (\textbf{f}) for either current-flow
direction.

Ideally, phase transitions are determined by specific heat. These measurements, however,
are time-consuming and may be hindered by torque, and, as will be described below, by sample 
fragility. Therefore, for the determination of phase transition fields and temperatures, 
we made use of magnetization and electrical resistivity data. Above, 
we found that \CeFe\ anomalies in the specific heat C$_\mathrm{p}$
are also found in susceptibility via Fisher's relation~\cite{Fisher62}
$\frac{\partial(\chi T)}{\partial T}$ and in the resistivity via $\frac{\partial\rho}{\partial T}$. 
Below, we will make use of a modified version of Fisher's relation, 
$\frac{\partial(M T)}{\partial T}$, and verify some of the results 
with $\frac{\partial\rho}{\partial T}$ from resistivity measurements, 
as well as with transition fields obtained from anomalies in $\frac{\partial M}{\partial H}$ 
obtained from magnetization isotherms~\cite{DeJongh74}.

\section{Pure \CeFe- results for \emph{H} $\bot$ \ca}
\subsection{Magnetization}

\label{Mperpc}

We did not observe any noticeable differences in magnetization for 
fields applied in different directions within the 
plane $\bot$ \ca, therefore, they will be presented as 
measured with \emph{H} $\bot$ \ca. Before we discuss these 
experiments, we address the fragility of the compound. 
It has been reported~\cite{Makihara03}
that at $\sim$ 5~K the crystallographic unit cell volume 
is larger in 5~T than in zero field, by $\sim$ 0.7 \%. 
At 1.8~K, a hysteretic metamagnetic 
transition, see below, takes place, presumably from a low-field, 
small-volume phase to a high-field larger-volume phase. 
In a magnetization experiment, at 1.8~K, on a plate-like sample,
the sample \emph{shattered} while the magnetic field 
(\emph{H} $\bot$ \ca) \emph{decreased} through this transition.

Although the data from the two types of crystal overlap, we found plate-like
samples to behave differently from blocky samples.
On blocky samples, we have been able to measure full-hysteresis 
magnetization isotherms at temperatures between 1.8~K and 300~K.
Visual inspection after the measurements revealed no obvious
damage. 
We do not know why plate-like samples appear more prone to shattering 
than the blocky samples. We may note, however, that the plate-like samples 
have a much larger surface-to-volume ratio than the blocky samples, and
the relative change of the \emph{a} and \emph{c} lattice parameters 
reportedly~\cite{Makihara03} is about equal. These effects can result 
in a relatively much larger relative change in surface area for the plate-like 
samples than for the blocky samples, when passing through the 
metamagnetic transition. 
Moreover, blocky samples may have more inclusions, which may prevent shattering. 

The plate-like samples, though, with their surface 
normal parallel to the \ca, were easy to (re)align, within  
2-3$^\circ$, and easier to clamp than the blocky samples. 
Because of their relative ease-of-use, we chose the plate-like 
samples for the detailed magnetization measurements, 
and developed a protocol, see below, that helps circumvent 
their fragility. For determining hysteresis behavior, we used blocky samples.

\begin{figure}[!ht]
\begin{center}
\includegraphics[bb=0 0 3.5in 3.5in,scale=0.9]{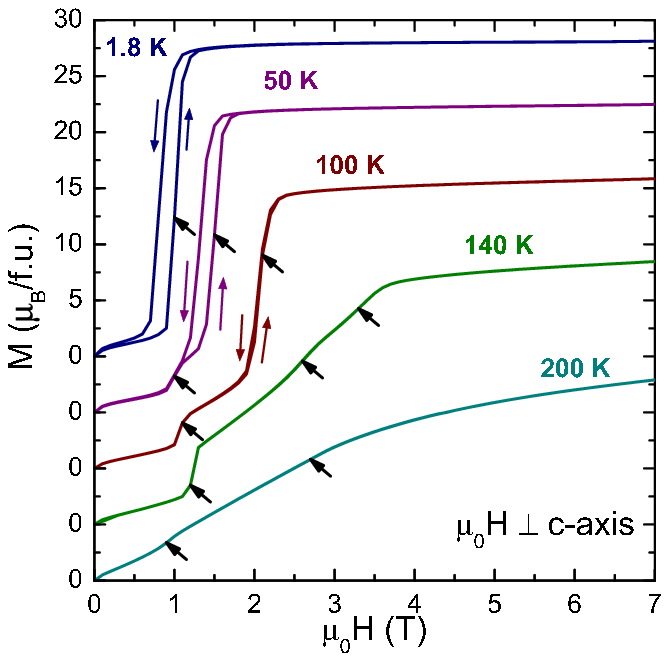}
\caption{Magnetization at various temperatures for \CeFe\ with the
magnetic field $\mu_0H$ $\bot$ \ca, measured with
increasing and decreasing fields with steps of 0.1~T. 
For clarity, the curves have been offset with respect to one another.
The black arrows indicate maxima in $\frac{\partial M}{\partial H}$ for increasing fields.}\label{hystplane}
\end{center}
\end{figure}

Fig.~\ref{hystplane} shows magnetization at various 
temperatures, measured on a blocky sample with both increasing and
decreasing field strengths up to 7~T, applied $\bot$ the \ca, with a 
step of 0.1~T between measurements.  
With increasing fields, the magnetization at 1.8~K is small and 
increases only slightly with increasing fields up to $\sim$ 1~T.
Then, around $\sim$ 1~T, the magnetization jumps to saturation and is
nearly field independent, reaching $\sim$ 27 $\mu_\mathrm{B}$/f.u. 
at 7~T. This jump in magnetization shows a hysteresis of about 0.2~T. 
The magnetization at 50 and 100~K also shows hysteretic jumps to
saturation, at $\sim$1.5 and $\sim$2.1~T, respectively. The width
of the hysteresis becomes smaller with increasing temperature. 
Besides the hysteretic metamagnetic transition in high fields, the
curves measured at 50 and 100~K both also show a 
step-like feature near 1~T with no detectable hysteresis,
indicative of another metamagnetic transition. 

At temperatures above $\sim$ 125~K, the lower zero-field transition 
temperature described in the previous section, magnetization 
isotherms are qualitatively different. 
At 140~K, besides a step in magnetization that occurs
here at $\sim$ 1.2~T,  the magnetization follows an s-shape increase with 
increasing fields with a bending point near 2.6~T, up to $\sim$3.3~T, where the
magnetization starts to saturate. Hysteresis is not observed.
At 200~K, still in the magnetically ordered phase, a weak feature is observed 
near $\sim$ 0.9~T, and a change in slope may be discerned near $\sim$ 2.7~T, 
indicating that the magnetization starts to saturate above this field strength. 

\begin{figure}[!ht]
\begin{center}
\includegraphics[bb=0 0 3.5in 3.5in,scale=0.9]{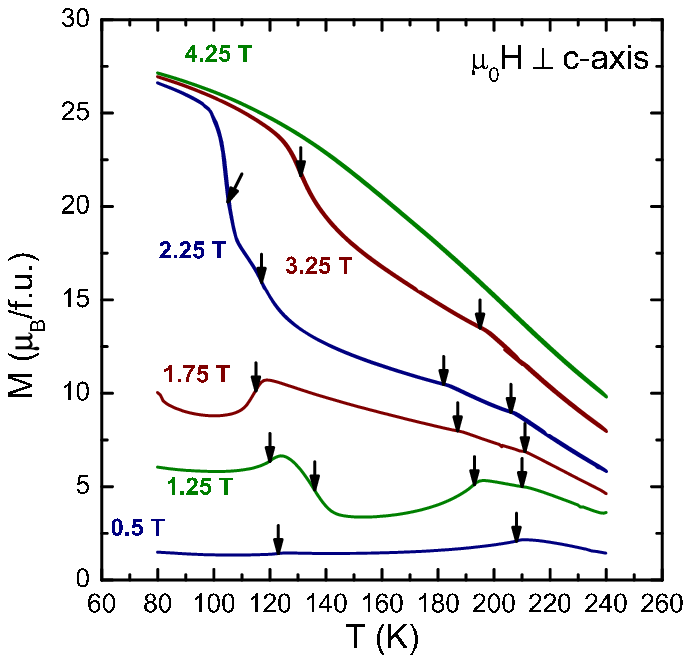}
\caption{Magnetization as a function of (decreasing) temperature
for fields ($\mu_0H$ $\bot$ \ca ) between 0.5 and 4.5
T. The arrows indicate maxima in $\frac{\partial (MT)}{\partial T}$.}\label{MTplanehiT}
\end{center}
\end{figure}

To avoid shattering of the plate-like samples, and still measure magnetization
at any field/temperature, we developed specific field and temperature protocols
strategies. As indicated above, the sample that
shattered, did so when it \emph{re}entered the low-temperature low-field phase through 
a hysteretic phase transition, such as those shown in Fig.~\ref{hystplane}.
Therefore, our measurements were performed so as to avoid such a
transition. For measurements of temperature-dependent magnetization, with temperatures 
reaching below $\sim$ 100~K, we used the following protocol.
In fields above $\sim$ 1~T, measurements were performed upon cooling from temperatures 
higher than the ordering temperature ($\sim $ 210~K). At low temperatures, 
the field was increased to above $\sim$ 4~T, which, as we will see below, is higher than  
any observed phase-transition field. After that, the sample was warmed to above 210~K, where 
the field was removed. Since the lowest field at which a hysteretic phase transition 
takes place is $\sim$ 1T, measurement in applied fields lower than this 
could be performed upon cooling as well as upon heating.

\begin{figure}[!ht]
\begin{center}
\includegraphics[bb=0 0 3.5in 7in,scale=0.9]{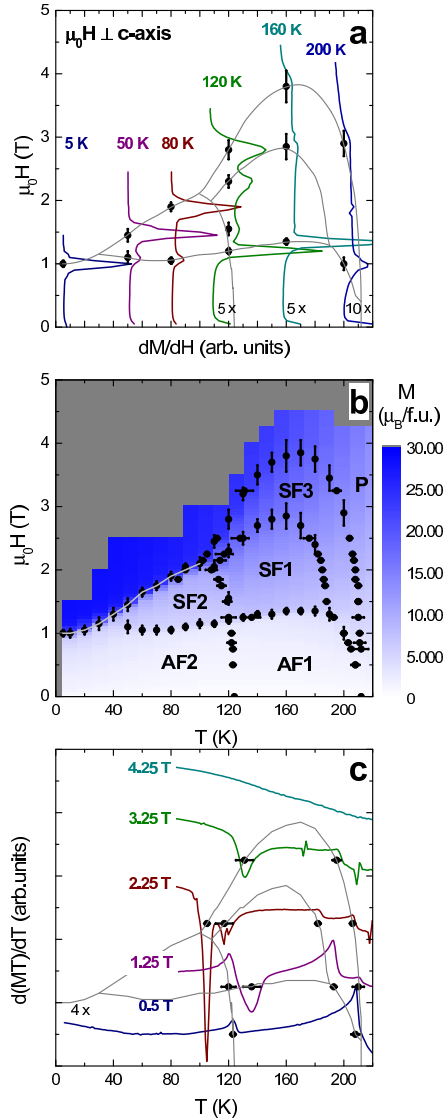}
\caption{Magnetic phase diagram for for \CeFe\ obtained with the
applied magnetic field \emph{H} $\bot$ \ca. 
\textbf{a} Traces of $\frac{\partial M}{\partial H}$ obtained from field-dependent magnetization. Peaks and width of peaks are indicated.  
\textbf{b} Contour plot of magnetization \emph{M}(\emph{H,T}) with peaks from \textbf{a} and \textbf{c} plotted on top. Labels are explained in the text. Dark grey indicates where no data was taken.
\textbf{c} Traces of $\frac{\partial (MT)}{\partial T}$ obtained from temperature-dependent magnetization. 
Peaks and width of peaks are indicated.}\label{planepd}
\end{center}
\end{figure}

Fig.~\ref{MTplanehiT} shows the field evolution of temperature 
dependent magnetization, 
measured on a plate-like sample, 
at temperatures between 240~K and 80~K, using this protocol. 
In a field of 4.25~T, the magnetization increases monotonically
with decreasing temperatures, without any clear anomalies. In lower fields, 
multiple anomalies occur, associated with magnetic phase transitions. 
Based upon the results of the previous section, where it was found that 
the peaks in the zero-field $\frac{\partial(\chi T)}{\partial T}$ appear very 
similar to the peaks in specific heat, we assume that peaks in 
$\frac{\partial (MT)}{\partial T}$ delineate magnetic phase boundaries. 
Anomalies determined from peaks in $\frac{\partial (MT)}{\partial T}$, 
are indicated by means of arrows in Fig.~\ref{MTplanehiT}. 
Traces of $\frac{\partial (MT)}{\partial T}$
are shown in Fig.~\ref{planepd}\textbf{c}, as well as the positions
of the peaks and their width, determined from half width 
at half maximum on the steepest flank.

Detailed measurements of field-dependent magnetization at fixed temperatures were 
performed using a similar protocol as above. For each temperature, 5~K, and 10~K 
up to 220~K with 10~K steps,
the sample was cooled in zero field, and the field was increased in steps of 
0.025~T up to the highest measuring field. This field was chosen such that 
the magnetization appeared clearly saturated.
After that, the field was increased up to above 4.5~T, without changing
the temperature, and then the temperature was increased to above 250~K, 
at which temperature the field was removed. 
The results of these measurements are shown in Fig~\ref{planepd}\textbf{b},
as a contour plot, that displays the magnetization as a function 
of applied field and temperature. 

As an often~\cite{DeJongh74,Stryjewski77,Morosan04} applied criterion for transition fields, 
according to mean-field theory (See e.g.\ ~\onlinecite{DeJongh74,Stryjewski77}), 
peaks in the differential susceptibility $\frac{\partial M}{\partial H}$, 
determined from magnetization isotherms, can be taken. 
Fig.~\ref{planepd}\textbf{a} shows several traces of $\frac{\partial M}{\partial H}$. 
The peaks, and their width, determined from the half width at half maximum on 
the steepest flank, are also indicated in Fig.~\ref{planepd}\textbf{a}. The differences between 
the peak positions indicated here and in Fig.~\ref{hystplane} can be ascribed 
to differences in demagnetizing fields, due to the different shapes 
of the samples.

Displayed together on top of the magnetization isotherms, in Fig.~\ref{planepd}\textbf{b},
the positions and the widths of the peaks obtained from both 
$\frac{\partial M}{\partial H}$ and from $\frac{\partial (MT)}{\partial T}$, appear 
to overlap. We therefore conclude that they indicate magnetic phase boundary lines, 
and thus outline a magnetic phase diagram. The phase boundary line that is associated 
with clear hysteresis is drawn in grey. For the other phase boundary lines, hysteresis was not
observed. The labels will be discussed below.

\begin{figure}[!ht]
\begin{center}
\includegraphics[bb=0 0 3.5in 5in,scale=0.9]{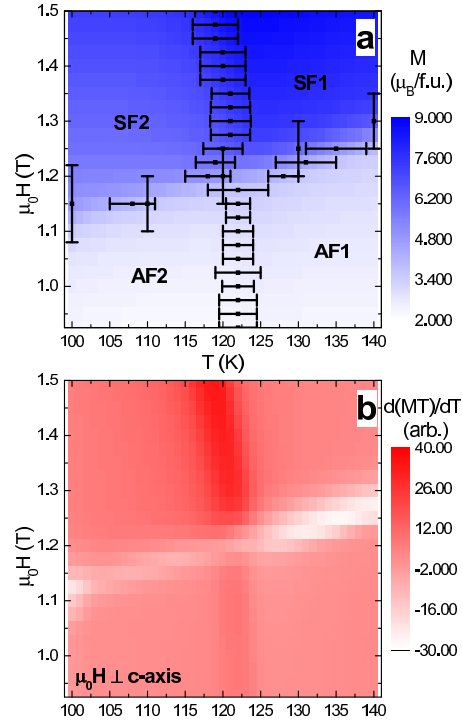}
\caption{Detail of \emph{H-T} phase diagram for \emph{H} $\bot$ \ca.
\textbf{a} Contour plot of magnetization \emph{M}(\emph{H,T}) with peaks determined from 
$\frac{\partial (MT)}{\partial T}$ (horizontal bars) and peaks determined from 
$\frac{\partial M}{\partial H}$ (vertical bars) plotted on top. 
\textbf{b} Contour plot of $\frac{\partial (MT)}{\partial T}$ determined from \textbf{a}.
}\label{X}
\end{center}
\end{figure}

To determine details of the apparent crossing of phase boundary lines 
at about 120~K and in about 1~T, we performed a more detailed experiment.
Temperature-dependent magnetization was measured upon cooling at 1~K steps 
between 140~K and 100~K. Fields ranged between 1.5~T and 0.925~T at 
0.025~T steps. The results of the measurements are displayed in Fig.~\ref{X}\textbf{a}, 
as a contour plot, that displays the magnetization as a function 
of applied field and temperature. In Fig.~\ref{X}\textbf{b}, we display 
the derivative $\frac{\partial (MT)}{\partial T}$ 
as a contour plot. The peak positions and their widths, obtained from 
this experiment are included in Fig.~\ref{X}\textbf{a}, as well as peak positions and  
widths obtained from the magnetization isotherms in Fig.~\ref{planepd}\textbf{b}.
From this it is clear that, within experimental resolution, there is a crossing
of phase boundary lines at $\sim$ 122~K in $\sim$ 1.2~T.
\subsection{Resistivity}
\label{resistivity}

\begin{figure}[!ht]
\begin{center}
\includegraphics[bb=0 0 3.5in 5in,scale=0.9]{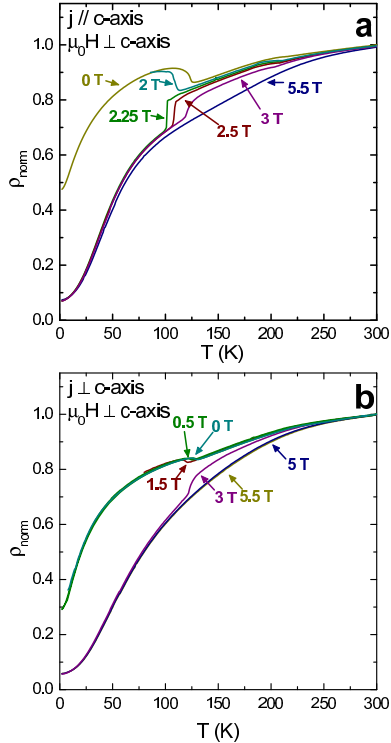}
\caption{
\textbf{a} Temperature-dependent normalized resistivity measured with \emph{j} // \ca\ in different \emph{H} $\bot$ \ca\ up to 5.5~T.
\textbf{b} as in \textbf{a}, but for \emph{j} $\bot$ \ca.
}\label{RHTfig}
\end{center}
\end{figure}

\begin{figure}[!ht]
\begin{center}
\includegraphics[bb=0 0 3.5in 5in,scale=0.9]{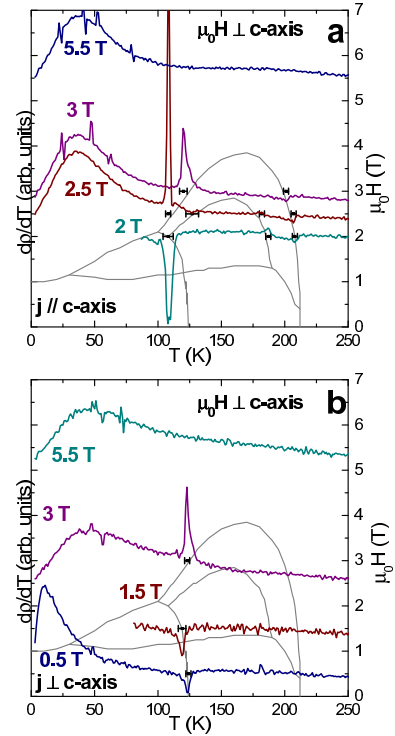}
\caption{
\textbf{a} Traces of $\frac{\partial \rho}{\partial T}$ for \emph{j} // \ca, plot on top of magnetic phase diagram for \emph{H} $\bot$ \ca.
\textbf{b} as in \textbf{a}, but for \emph{j} $\bot$ \ca.
}\label{RHTderiv}
\end{center}
\end{figure}

We determined temperature-dependent resistivity for the current \emph{j} flowing // \ca\ as well as for \emph{j} $\bot$ \ca\ in various applied fields \emph{H} $\bot$ \ca\ up to 5.5~T.  After the zero-field measurements, we performed measurements in applied fields. With the fragility of the material in mind, we measured, starting at 300~K, with both decreasing and increasing temperatures. In this way, we verified that the samples remained intact, which was obvious if, at a given temperature, the resistance measured upon heating was found to be the same as the resistance measured upon cooling. The measurements with \emph{j} // \ca\ were performed at gradually lower fields, starting at 5.5~T. Down to 2.25~T, there was no significant change in the resistance determined with increasing or with decreasing temperatures. At 2~T, however, upon cooling below $\sim$ 86~K, the sample resistance became unstable: it started to fluctuate, and the resistance upon warming was higher than upon cooling. Note that at this temperature and field the phase diagram, Fig.~\ref{planepd}, shows a hysteretic transition. Although visual inspection did not reveal any obvious damage, the sample was no longer usable. For the magnetization measurements, we were able to prevent the samples from shattering by crossing the hysteretic phase boundary only with increasing field
or decreasing temperature. Apparently, for the resistivity measurements, such a hysteretic phase transition has to be completely avoided. For this reason the lowest temperature for the measurement in 1.5~T  with \emph{j} $\bot$ \ca\ was 80~K.
Besides these, for the other applied-field values, measurements were performed every 1~K between 300~K and 2~K. 

The results are shown in Fig.~\ref{RHTfig}\textbf{a}  for \emph{j} // \ca, and in Fig.~\ref{RHTfig}\textbf{b} for \emph{j} $\bot$ \ca. For completeness, both Fig.~\ref{RHTfig}\textbf{a} and \textbf{b} include the zero-field results from Fig~\ref{zerofielddata}. Figs.~\ref{RHTderiv}\textbf{a} and \textbf{b} shows traces of $\frac{\partial \rho}{\partial T}$ plot over the phase diagram from Fig.~\ref{planepd}. Also included in Fig.~\ref{RHTderiv} are bars that indicate peak positions and widths in $\frac{\partial \rho}{\partial T}$. Note that the peaks in $\frac{\partial \rho}{\partial T}$ coincide well with the previously determined phase diagram.

The anomalous increase in resistivity with decreasing temperature, effects we ascribed above to the opening of an anisotropic superzone gap, observed in zero field for both current directions, occurs also in applied fields for both current directions. We observed it for fields of 2~T (\emph{j} // \ca), 1.5~T, 0.5~T (\emph{j} $\bot$ \ca). Note that in zero field and in 0.5~T, the crossed boundary lies between phases marked as AF1 and AF2, in Figs.~\ref{planepd} and \ref{X}, whereas for 1.5~T and 2~T, the crossed boundary lies between the phases marked as SF and SF2.
 
Generally, for both employed current directions, the high-field resistivity at low temperatures is low, whereas for low fields it is high. The residual resistance ratio in 5.5~T, about 14 and 17, for \emph{j} // \ca\ and \emph{j} $\bot$ \ca, respectively, is much higher than in zero field. Since, as already indicated above, in Sec.~\ref{Zero-fielddata}, the resistivity for \emph{j} $\bot$ \ca\ is hardly affected by superzone-gap effects, whereas its low-temperature resistivity is much lower in 5.5~T than in zero-field, the superzone gap does not determine the large magnetoresistance, as was previously noted~\cite{Janssen97}. Assuming the resistivity is determined as 
$\rho=\rho_0+\rho_\mathrm{lattice}+\rho_\mathrm{mag}$, and that the residual resistivity, $\rho_0$, and the lattice contribution $\rho_\mathrm{lattice}$ are the same for the low-temperature AF2 or P phase, we consider the large magnetoresistance mainly caused by a strong change in magnetic scattering $\rho_\mathrm{mag}$. 

For \emph{j} $\bot$ \ca\ the anomalies associated with phase transitions are weaker than for \emph{j} // \ca. The ordering transition between the paramagnetic phase P and AF1 (for 0.5~T) or between P and SF3 (for 1.5 and 3~T) is not observed as an anomaly for \emph{j} $\bot$ \ca, nor is the transition between SF3 and SF1 (for 1.5~T). For \emph{j} $\bot$ \ca, the transition between SF3 and P produces a clear anomaly. For \emph{j} // \ca, weak anomalies mark the phase boundaries between P and SF3 (at 2~T, 2.5~T and 3~T). For the purpose of this paper, we do not consider the 'break' in resistivity around 175~K,  for \emph{j} // \ca\ in 5.5~T, resulting in a very broad 'peak' in $\frac{\partial \rho}{\partial T}$, due to a phase transition.

Although the presented dataset is limited, we have  been able to measure resistivity in every (\emph{H,T}) phase in the phase diagram of Fig.~\ref{planepd}. Our results indicate that the previously observed large magnetoresistance~\cite{Janssen97}, is mainly due to a large (magnetic) scattering of conduction electrons in the low-temperature AF1 phase, which is strongly reduced in the field-induced P state. Here, the superzone gap is associated with the phase-boundary line that separates AF1 and SF1 (above 124~K), from AF2 and SF2. 

\section{Pure \CeFe- magnetization results for \emph{H} // \ca}

\label{Mparc}

In this section we discuss magnetization for the 
field applied // \ca, resulting in an \emph{H-T} magnetic phase diagram for this 
applied-field direction. As in Sec.~\ref{Mperpc}, we used blocky samples for determining hysteresis behavior, and
we used plate-like samples for detailed magnetization measurements. 

\begin{figure}[!ht]
\begin{center}
\includegraphics[bb=0 0 3.5in 3.5in,scale=0.9]{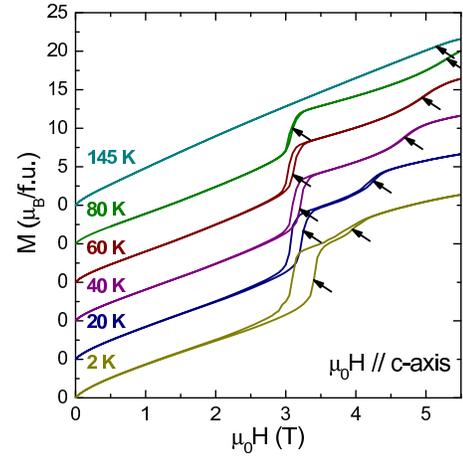}
\caption{Magnetization at various temperatures for \CeFe\ with the
magnetic field $\mu_0H$ // \ca, measured with
increasing and decreasing fields with steps of 0.1~T. 
For clarity, the curves have been offset with respect to one another.
The black arrows indicate maxima in $\frac{\partial M}{\partial H}$ for increasing fields.}\label{hystaxis}
\end{center}
\end{figure}

Fig.~\ref{hystaxis} shows magnetization at various 
temperatures up to 145~K, measured on a blocky sample with both 
increasing and decreasing field strengths up to 5.5~T, 
applied // \ca, with a step of 0.1~T between measurements.  
With increasing fields, the magnetization at 2~K is small and 
increases slightly with increasing fields up to $\sim$ 3.2~T.
Then, around $\sim$ 3.2~T, a metamagnetic transition takes place, 
followed by a second metamagnetic transition at $\sim 3.9~T$. Both these 
transitions show hysteresis. At 5.5~T the magnetization is not saturated, 
and it reaches a value of $\sim$ 26 $\mu_\mathrm{B}$/f.u.\, lower than the 
magnetization observed at 1.5~T for \emph{H} $\bot$ \ca,
in Fig.~\ref{hystplane}.  The lower-field metamagnetic transition occurs 
at temperatures below $\sim$ 125~K, at which a zero-field transition occurs.    
The (increasing) field, at which this metamagnetic transition occurs, 
decreases slightly up to 80~K, and at the same time the hysteresis gradually 
becomes smaller. The field at which the higher-field transition occurs, 
increases with increasing temperature, at least up to 80~K. The hysteresis 
associated with this transition is observed at 20~K, but not at 40~K, 
nor at higher temperatures. At 145~K, the magnetization increases smoothly, 
and without anomalies, until it approaches saturation near 5.5~T.  

\begin{figure}[!ht]
\begin{center}
\includegraphics[bb=0 0 3.5in 3.5in,scale=0.9]{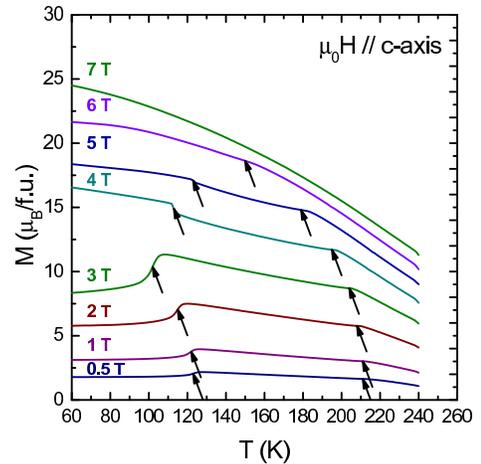}
\caption{Magnetization as a function of (decreasing) temperature
for fields ($\mu_0H$ // \ca ) between 0.5 and 7
T. The arrows indicate maxima in $\frac{\partial (MT)}{\partial T}$.}\label{MTaxis}
\end{center}
\end{figure}

\begin{figure}[!ht]
\begin{center}\includegraphics[bb=0 0 3.5in 7in,scale=0.9]{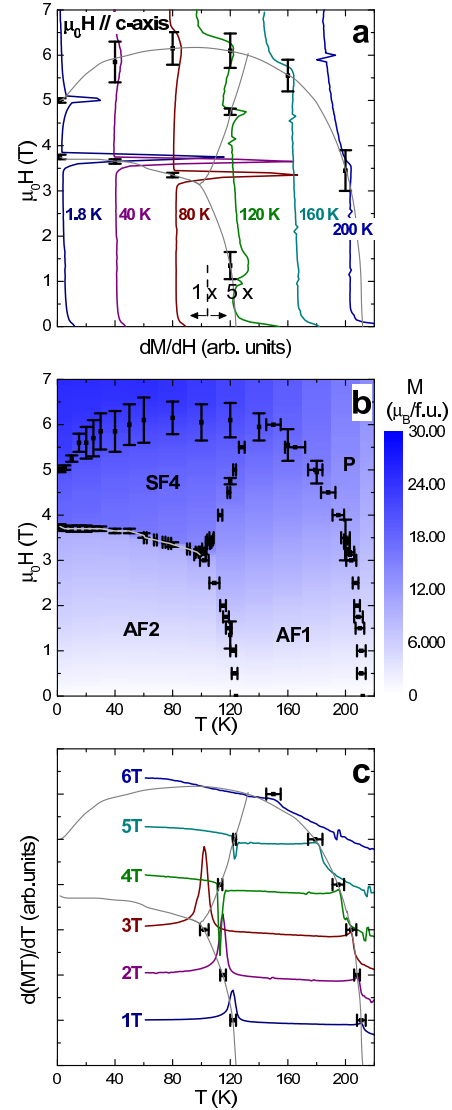}
\caption{Magnetic phase diagram for for \CeFe\ obtained with the
applied magnetic field \emph{H} // \ca. 
\textbf{a} Traces of $\frac{\partial M}{\partial H}$ obtained from field-dependent magnetization. Peaks and width of peaks are indicated.  
\textbf{b} Contour plot of magnetization \emph{M} (\emph{H,T}) with peaks from \textbf{a} and \textbf{c} plotted on top. Labels are explained in the text. 
\textbf{c} Traces of $\frac{\partial (MT)}{\partial T}$ obtained from temperature-dependent magnetization. 
Peaks and width of peaks are indicated.}\label{axispd}
\end{center}
\end{figure}

Fig.~\ref{MTaxis} shows the field evolution of temperature 
dependent magnetization, measured on a plate-like sample, 
at temperatures between 240~K and 60~K. 
A similar strategy as described above in Sec.~\ref{Mperpc} was used. 
We measured upon cooling, and at 60~K, we increased the field up to 7~T, 
and heated the sample to  $\sim$ 240~K, before we removed the field. 
In a field of 7~T, the magnetization increases uniformly 
with decreasing temperatures, without clear anomalies. In lower fields 
multiple anomalies occur, associated with magnetic phase transitions. 
Based upon the results of the previous sections, we assume that peaks in 
$\frac{\partial (MT)}{\partial T}$ in applied fields denote magnetic phase boundaries. 
Anomalies determined from peaks in $\frac{\partial (MT)}{\partial T}$, 
are indicated by means of arrows in Fig.~\ref{MTaxis}. 
Traces of $\frac{\partial (MT)}{\partial T}$
are shown in Fig.~\ref{axispd}\textbf{c}, as well as the positions
of the peaks and their width, determined from half width 
at half maximum on the steepest flank.

Detailed measurements of field-dependent magnetization at fixed temperatures were 
performed using a similar protocol as above. For each temperature, 5~K, and 10~K 
up to 220~K with a 20~K step,
the sample was cooled in zero field, and the field was increased in steps of 
0.05~T up to 7~T. Then the temperature was increased to above 250~K, 
at which temperature the field was removed. The results of these measurements
are shown in Fig~\ref{axispd}\textbf{b}, as a contour plot, that displays the magnetization 
as a function of applied field and temperature. 

The anomalies observed in these magnetization isotherms produced 
peaks in the derivatives $\frac{\partial M}{\partial H}$. 
Fig.~\ref{axispd}\textbf{a} shows several traces of $\frac{\partial M}{\partial H}$. 
The peaks, and their width, determined from the half width at half maximum on 
the steepest flank, are also indicated in Fig.~\ref{axispd}\textbf{a}
The differences between the peak positions indicated here and in Fig.~\ref{hystaxis}
can be ascribed to differences in demagnetizing fields, due to the different shapes 
of the samples.

As for Fig.~\ref{planepd}, in Fig.~\ref{axispd}\textbf{b}
the positions and the widths of the peaks obtained from both 
$\frac{\partial M}{\partial H}$ and from $\frac{\partial (MT)}{\partial T}$, overlap. 
We therefore conclude that they indicate magnetic phase boundary lines, 
and thus outline a magnetic phase diagram. Phase boundary lines 
associated with hysteresis are indicated with gray lines. For the other phase boundary lines, 
hysteresis was not observed. The labels will be discussed below.

\begin{figure}[!ht]
\begin{center}\includegraphics[bb=0 0 3.5in 5in,scale=0.9]{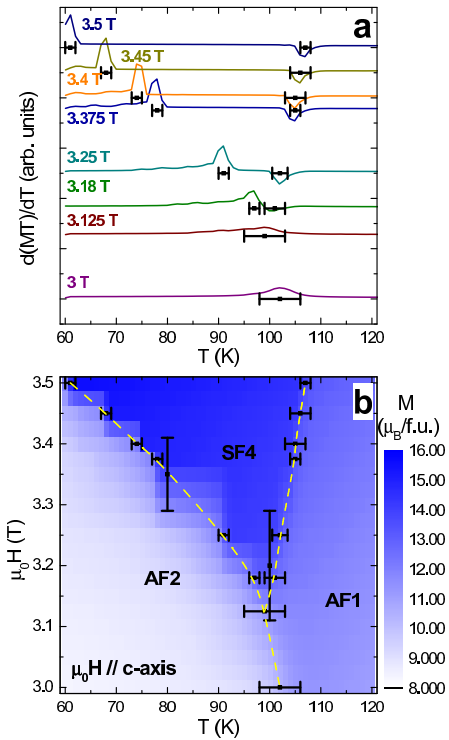}
\caption{Detail of \emph{H-T} phase diagram for for \emph{H} // \ca. 
\textbf{a} Traces of $\frac{\partial (MT)}{\partial T}$ obtained from temperature-dependent magnetization. 
Peaks and width of peaks are indicated as horizontal bars.
\textbf{b} Contour plot of magnetization \emph{M} (\emph{H,T}), obtained via interpolation, with peaks determined from 
$\frac{\partial (MT)}{\partial T}$ (horizontal bars) and peaks determined from 
$\frac{\partial M}{\partial H}$ (vertical bars) plotted on top. 
}\label{Y}
\end{center}
\end{figure}

Fig.~\ref{Y} shows details of the apparent splitting of phase boundary lines 
at about 100~K and in about 3.2~T.
Temperature-dependent magnetization was measured upon cooling at 1~K steps 
between 140~K and 60~K. Fields ranged between 3.5~T and 3.0~T at irregular intervals. 
The results of the measurements are displayed in Fig.~\ref{Y}\textbf{b}, 
as a contour plot, that displays the magnetization as a function 
of applied field and temperature. The contour plot shows field steps of 0.025~T, 
values for magnetization in fields between the measurement fields were obtained by interpolation.
In Fig.~\ref{Y}\textbf{a}, we display traces of the derivative $\frac{\partial (MT)}{\partial T}$. 
The peak positions and their widths, obtained from 
this experiment are included in Fig.~\ref{Y}\textbf{a}. Fig.~\ref{Y}\textbf{b} also contains these peak positions and their widths,
as well as widths obtained from the magnetization isotherms in Fig.~\ref{axispd}\textbf{b}.
From this it is clear, that within experimental resolution, there is a splitting 
of phase boundary lines at $\sim$ 100~K in $\sim$ 3.15~T.

\section{Pure \CeFe- angle-dependent magnetization}

\begin{figure}[!ht]
\begin{center}\includegraphics[bb=0 0 3.5in 5in,scale=0.9]{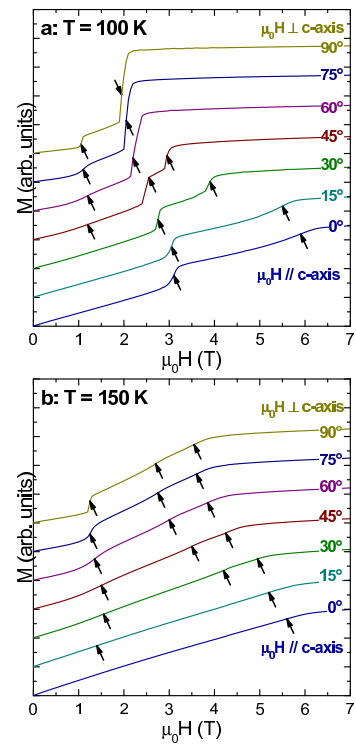}
\caption{Evolution of magnetization isotherms, at 100~K (\textbf{a}, and at 150~K (\textbf{b}), measured at angles with the \ca, with a 15$^\circ$ step, from // \ca, to $\bot$ \ca. For clarity the curves are offset with respect to one another. The arrows indicate extrema in $\frac{\partial M}{\partial H}$.   
\label{magnrot}
}\end{center}
\end{figure}

\begin{figure}[!ht]
\begin{center}\includegraphics[bb=0 0 3.5in 5in,scale=0.9]{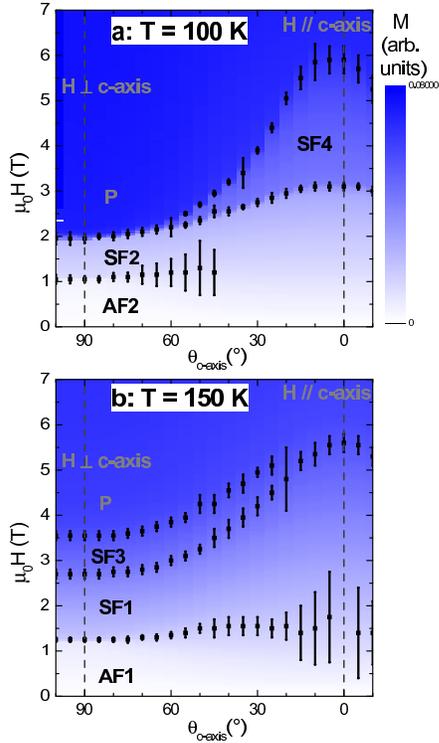}
\caption{Magnetic phase diagrams determined at 100~K and 150~K for fields applied at various angles $\theta$ with the \ca.
\textbf{a} Contour plot of magnetization \emph{M}(\emph{H},$\theta$) at 100~K with peaks determined from 
$\frac{\partial M}{\partial H}$ (vertical bars) plotted on top.
Labels are explained in the text.
\textbf{b} Contour plot of magnetization \emph{M} (\emph{H},$\theta$) at 150~K with peaks determined from 
$\frac{\partial M}{\partial H}$ (vertical bars) plotted on top.
Labels are explained in the text.
}\label{rotator}
\end{center}
\end{figure}

Magnetization for fields applied at different angles $\theta$ with the \ca, was measured on a plate-like sample (mass $<$ 0.5 mg), mounted on a rotator.
We measured over a range of 130$^{\circ}$ with 5$^{\circ}$ steps, overshooting both \emph{H} // \ca\ and \emph{H} $\bot$ \ca. As above, to preserve the sample, we used a protocol. In zero field, and at $\sim$ 240~K, $\theta$ was set. Then the sample was cooled to the measurement temperature, at which the magnetization was measured with increasing fields, every 0.05~T up to 7~T. After that, the sample was heated up, in 7~T, to $\sim$ 240~K, at which temperature the field was removed, and a new $\theta$ was set.    

The evolution of the magnetization, as $\theta$ is varied, for both \emph{T} = 100~K and \emph{T} = 150~K, is repectively shown in Fig.~\ref{magnrot}\textbf{a} and in Fig.~\ref{magnrot}\textbf{b}. Arrows indicate extrema found in $\frac{\partial M}{\partial H}$, note that some extrema are more clearly visible than others (cf.\ Figs.~\ref{hystplane} and \ref{hystaxis}). 
For \emph{H} // \ca, at 150K, th magnetization increases smoothly, up to about 5.5~T, where the magnetization starts to saturate. No s-shapes marking phase transitions were observed for \emph{H} // \ca. As $\theta$ increases, near 1-2~T an s-shape appears, that gradually develops into the step-like transition observed for \emph{H} $\bot$ \ca. The transition that marks the field where the magnetization starts to saturate, gradually decreases with increasing $\theta$. 
Furthermore, with $\theta$ increasing above $\sim$ 30 $^\circ$, an s-shape appears at fields slightly lower than the saturation field.   

Also at 100~K, with increasing $\theta$, the field at which the magnetization starts to saturate decreases gradually. However, here it develops from a smooth, slightly s-shaped, non-hysteretic, transition into a very sharp and hysteretic transition, see Figs.~\ref{hystaxis} and \ref{hystplane}, respectively. With increasing $\theta$, the sharp (and hysteretic) transition observed for \emph{H} // \ca\ near 3~T shifts to a slightly lower field, and for  $\theta > 45^\circ$, appears to merge with the jump-like transition to saturation observed for \emph{H} $\bot$ \ca. Furthermore, the step-like transition near 1.5~T, observed for \emph{H} $\bot$ \ca\ develops from an s-shape that first appears for $\theta > 30^\circ$. 

Contour maps of \emph{M}($\theta$,\emph{H}) measured at 100 (\textbf{a}) and 150~K (\textbf{b}), which is below and above the 125~K zero-field transition, respectively, are displayed in Fig.~\ref{rotator}. Vertical bars denote peaks in $\frac{\partial M}{\partial H}$ and their widths obtained as above in Secs.~\ref{Mperpc} and ~\ref{Mparc}. Note that, especially for the less-visible transitions, the peak widths may be slightly underestimated (cf.\ Figs.~\ref{planepd} and \ref{axispd}). For Fig.~\ref{rotator}\textbf{a}, measured at 100~K, for \emph{H} $\bot$ \ca, transitions were found near $\sim$ 1~T and near $\sim$ 2~T, just as in Fig.~\ref{planepd}. At the same temperature, for \emph{H} // \ca, transitions were found near $\sim$ 3~T and near $\sim$ 5.7~T, just as in Fig.~\ref{axispd}. Fig.~\ref{rotator}\textbf{b}, measured at 150~K, compares similarly to Figs.~\ref{planepd} and ~\ref{axispd}. The labels are discussed in Sec.~\ref{discussion}.

\section{Ta-doped \CeFe- Magnetization results for \emph{H} $\bot$ \ca}

\begin{figure}[!ht]
\begin{center}\includegraphics[bb=0 0 3.5in 7in,scale=0.9]{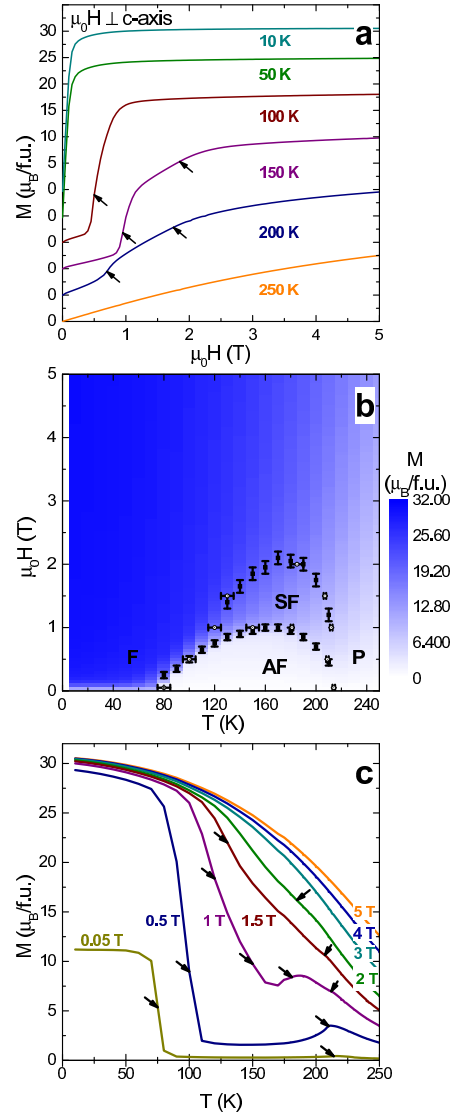}
\caption{
\textbf{a} Magnetization at various temperatures for Ta-doped \CeFe\ with the
magnetic field $\mu_0H$ $\bot$ \ca, measured with
increasing fields with steps of 0.1~T. 
For clarity, the curves have been offset with respect to one another.
The black arrows indicate maxima in $\frac{\partial M}{\partial H}$.
\textbf{b} Contour plot of magnetization \emph{M} (\emph{H,T}) with peaks from $\frac{\partial M}{\partial H}$ and $\frac{\partial (MT)}{\partial T}$
plotted on top. 
Labels are explained in the text.
\textbf{c} Temperature-dependent magnetization in various fields up to 5~T. Arrows indicate phase-transition temperatures.
}\label{Tadoped}
\end{center}
\end{figure}

In order to compare undoped \CeFe\ to Ta-doped \CeFe, we used the Ta-doped \CeFe\ crystal with nominally 0.5 \% Ta, that was also used for the EPMA experiment. We chose to measure magnetization with the field \emph{H} $\bot$ \ca, because, as will become clear below, this is the most salient applied-field direction. 
Fig.~\ref{Tadoped} shows a summary of these measurements. Fig.~\ref{Tadoped}\textbf{c} shows temperature-dependent magnetization curves, measured in 0.05, 0.5, 1, 1.5, 2, 3, 4, and 5~T, at (decreasing) temperatures between 250~K and 5~K. In curves of $\frac{\partial (MT)}{\partial T}$ (not shown), extrema were found, which are indicated by arrows in Fig.~\ref{Tadoped}\textbf{c}. In fields of 3~T and higher, no sharp anomalies were observed, similar to undoped \CeFe\ for \emph{H} $\bot$ \ca\ in 4.25~T (see Fig.~\ref{MTplanehiT}). In the lower field of 2~T, anomalies appear, marking phase transitions. In contrast to undoped \CeFe, a large magnetization is found below $\sim$ 75~K, in all applied fields of 0.05~T and higher, indicating that our sample of Ta-doped \CeFe\ is ferromagnetic at low temperatures. The multiple anomalies in fields higher than 0.05~T, especially in 1~T, hint at a complex phase diagram. The low-field 0.05~T results are consistent with two ordering transitions, antiferromagnetic at \TN\ = 214(1), and ferromagnetic at \TO\ = 75(5)~K. 
Results of (increasing) field-dependent magnetization at various temperatures between 10~K and 200~K are displayed in Fig.~\ref{Tadoped}\textbf{a}. 
At 10~K and at 50~K the magnetization shows ferromagnetic behavior, and no metamagnetic transitions are observed. The saturation magnetization of $\sim$ 31 $\mu_\mathrm{B}$/f.u.\ at 10~K is in good agreement with results published for Ta-crucible grown \CeFe~\cite{Makihara04}.
At 100~K, 150~K, and 200~K, metamagnetic transitions were observed as maxima in $\frac{\partial M}{\partial H}$, at positions which are indicated by arrows. At 100~K, only one anomaly was observed, whereas at 150 and 200~K two anomalies were observed, indicating the existence of a metamagnetic phase at these temperatures, and consistent with the multiple anomalies found in temperature-dependent magnetization. 
Fig.~\ref{Tadoped}\textbf{b} shows a contour map of \emph{M}(\emph{H,T}), and phase boundaries obtained from $\frac{\partial M}{\partial H}$ (vertical bars) and $\frac{\partial (MT)}{\partial T}$ (horizontal bars). The labels are analogous to those given for pure \CeFe, but here is only one zero-field AF phase.

\section{Discussion}
\label{discussion}

The detailed study of the magnetic phase diagrams of undoped and of Ta-doped \CeFe\ provides insight into the competition between ferro- and antiferromagnetic interactions in these compounds.
In a magnetic phase diagram, one can find a phase-boundary line starting, in zero field, at the N\'{e}el temperature. It separates the paramagnetic phase from the magnetically ordered phase. Generally, its critical magnetic field strength increases with decreasing temperature~\cite{DeJongh74}.
However, for Ta-doped \CeFe, the ground state is not antiferromagnetic but ferromagnetic. In such a case, the antiferromagnetic transition fields may form a 'dome' in \emph{H-T}, as predicted in Ref.~\onlinecite{Moriya77}. Such a dome is indeed observed in the \emph{H-T} phase diagram of Ta-doped \CeFe, which in zero fields orders AF at the N\'{e}el temperature \TN\ $\sim$ 214~K, and ferromagnetic (F) at \TO $\sim$ 75~K. 
Furthermore, besides the AF dome, there is an additional dome, enclosing a metamagnetic spin-flop-like phase, SF. 
This dome has a maximum field of $\sim$ 2~T near 170~K. 

Undoped \CeFe\ also order AF, at \TN\ $\sim$ 208~K. Above \TT\ $\sim$ 124~K, also the \emph{H-T} phase diagram for \emph{H} $\bot$ \ca\, Fig~\ref{planepd}, shows an AF dome. Besides this, at there are two more domes, enclosing metamagnetic spin-flop-like phases, SF1 and SF3. The dome enveloping SF3 has a maximum field of $\sim$ 4~T at $\sim$ 170~K. However, the boundary of this high-temperature SF3 dome does not go down to zero field at some lower temperature, instead it converges with another, low-temperature, dome, which is connected to \TT\ at $\sim$ 124~K, which is, as reported elsewhere in this volume~\cite{Kreyssig06}, associated with a crystallographic phase transition. This low-temperature dome also encloses a metamagnetic spin-flop-like phase, labeled SF2. The result is that undoped \CeFe\ remains antiferromagnetic below \TT\ (AF2). 

As demonstrated in Fig.~\ref{X}, there is a crossing of phase boundary lines at $\sim$122~K in $\sim$1.2~T. According to Landau~\cite{Landau37a}, for continuous phase transitions such crossings are only possible if the crossing phase boundary lines represent different order parameters. Although from mean-field descriptions it may be concluded that spin-flop transitions are first order~\cite{DeJongh74}, we observed no hysteresis in the measurements of magnetization around the phase boundary line separating the low-field phases AF1 and AF2 from the spin-flop-like phases SF1 and SF2, respectively. The phase boundary that starts at \TT\ in zero fields is characterized by a superzone-gap-like feature in electrical resistivity. 
The superzone-gap-like behavior was observed for the phase transition from AF1 to AF2, and between SF1 and SF2. We therefore assume that the field-dependent superzone-gap-like behavior is connected to the observed zero-field crystallographic phase transition~\cite{Kreyssig06}. 

The low-temperature dome that starts at \TT\ in zero field ends in about 1~T at 0~K. The boundaries of the low-temperature dome indicated by a grey line in Fig.~\ref{planepd} are due to hysteretic phase transitions. The boundaries of the high-temperature dome do not show hysteresis. The high-temperature dome is associated with an antiferromagnetic phase transition, and the low-temperature dome with a crystallographic phase transition. According to Landau~\cite{Landau37a}, a convergence of two phase-boundary lines of continuous phase transitions may produce a line of first-order phase transitions, under certain conditions~\cite{Yip91}. In this picture, the grey line of hysteretic first-order phase transitions constitutes a combined crystallographic and antiferromagnetic phase transition.

Now we turn to the \emph{H-T} phase diagram \emph{H} // \ca\ for undoped \CeFe, Fig.~\ref{axispd}. 
At the N\'{e}el temperature, a phase boundary line between AF1 and P starts. It is connected to a phase boundary line enveloping SF4, and a phase boundary line between SF4 and AF1. 
The boundary between SF4 and P becomes hysteretic below $\sim20$~K. If the s-shape magnetization just below 140~K is due to a continuous phase transition, then this phase boundary line may have a tricritical~\cite{DeJongh74, Stryjewski77} point near 20~K, see Fig~\ref{hystaxis}.  

Connected to \TT\ is a phase-boundary line, between AF1 and AF2, that is associated with a crystallographic phase transition~\cite{Kreyssig06}, this line is connected to the hysteretic phase-boundary line between AF2 and SF4, and to the phase-boundary line between AF1 and SF4. Obviously, the hysteretic phase boundary line between AF2 and SF4 represents first-order phase transitions. Then, following thermodynamic arguments~\cite{Landau37a,Yip91}, the phase-boundary line between AF1 and AF2, and the phase-boundary line between AF1 and SF4 may represent continuous phase transitions at this point. 

Comparing the magnetization behavior near the \TT-line in fields below $\sim$2.5~T, it appears that the magnetization at lower temperatures is lower than at higher temperatures. Thus the crystallographic phase transition~\cite{Kreyssig06} affects the magnetic properties for this applied field direction. 
The situation is different for higher fields, because at a given field above $\sim$3.5~T, the magnetization in the low-temperature phase SF4 is higher than in the neighboring phase AF1. This suggests that the phase boundary line between AF1 and SF4 is associated with the occurrence of a  low-temperature field-stabilized phase. 

At 150~K, well above \TT, the antiferromagnetic phase transition is observed for both \emph{H} // \ca\ and for \emph{H} $\bot$ \ca\, near 3.5~T and 6~T respectively, whereas spin-flop-like phase-transitions are only observed for \emph{H} // \ca. 
The angle-dependent magnetization experiment, Fig.~\ref{rotator}\textbf{b} shows how the antiferromagnetic transition and the transitions between AF1 and SF1, and between SF1 and SF3, respectively, evolve, as the angle between the applied field and the \ca\ is changed.  
In simple cases~\cite{DeJongh74}, spin-flop transitions are observed for weakly anisotropic antiferromagnets, when the field is applied parallel to the ordered moment direction. In this picture, the angle-dependent results measured at 150~K indicate that the moments in \CeFe\ order in the plane perpendicular to the \ca. Moreover, the angle-dependent behavior of the antiferromagnetic phase-transition field, which is higher for \emph{H} // \ca\ than for \emph{H} $\bot$ \ca, is also indicative of a magnetic anisotropy in favor of the plane $\bot$ \ca.
Also at 100~K, well below \TT, the angle-dependence of the antiferromagnetic phase-transition field, higher for \emph{H} // \ca\ than for \emph{H} $\bot$ \ca, indicates a magnetic anisotropy in favor of the plane $\bot$ \ca. 
The presence of the non-hysteretic spin-flop-like transition between AF2 and SF2, for \emph{H} $\bot$ \ca, and its absence for \emph{H} // \ca, is also an indication that at 100~K the ordered magnetic moments lie in the plane $\bot$ \ca.
The angle-dependent magnetization furthermore indicates that SF4, observed for \emph{H} // \ca, is not connected to any phase that can be observed for \emph{H} $\bot$ \ca. Therefore the angle-dependent results are consistent with the interpretations that the hysteretic AF2-SF4 phase-boundary line includes the crystallographic phase transition, and that the phase-boundary line enveloping SF4 is related to the magnetic anisotropy, as also observed at 150~K. 

We think the \emph{H-T} phase diagrams of Ta-doped \CeFe\ and of undoped \CeFe\ are manifestations of competing antiferromagnetic and ferromagnetic exchange interactions. The occurrence of magnetic phase transitions in itinerant-electron systems with one type of ferromagnetism and one type of antiferromagnetism in competition, has been treated theoretically by Moriya and Usami~\cite{Moriya77}. 
They constructed possible phase diagrams, one of which is very similar to the Ta-doped \CeFe\ for \emph{H} $\bot$ \ca. This predicted phase diagram has a low-temperature ferromagnetic state, an antiferromagnetic phase at intermediate temperatures, and a paramagnetic state at high temperatures. 
In non-zero applied fields, the antiferromagnetic phase is enveloped by a dome.

Although similar, the phase diagram for Ta-doped \CeFe, Fig.~\ref{Tadoped} is more complex than the one described by Moriya and Usami, because here the AF dome shows another dome on top of it, indicated as SF. 
In the work of Moriya and Usami, calculations are made for two competing and interacting magnetic modes, a uniform magnetization \emph{M}$_0$, and a staggered magnetization \emph{M}$_Q$, for the competing ferromagnetic state with wavevector 0 and antiferromagnetic state with wave vector \emph{Q}, respectively. The free energy was expanded up to the fourth order in these magnetic modes, without considering magnetic anisotropy. 
It therefore seems that the model can be adjusted by including higher-order terms in magnetization, by including magnetic anisotropy or by introducing more than two competing types of exchange interactions. Although the, isotropic, Moriya-Usami model has not been extended with an in-plane easy-plane anisotropy, no in-plane anisotropy has been observed for \CeFe, and easy-plane systems, with moments confined to the easy plane are expected to behave similarly to isotropic systems. Higher order terms in magnetization are not expected to qualitatively change the possible phase diagrams~\cite{Moriya77}.
We think that an extension of the theory by possibly including more than two competing and interacting magnetic modes may result in a phase diagram similar to the one we observed for Ta-doped \CeFe. This seems plausible, in view of the complexity of the crystal structure, with 5 different crystallographic sites.  Moreover, electronic-structure calculations on related \YFe\ produced different exchange interaction parameters of both antiferromagnetic and ferromagnetic type, between these sites~\cite{Sabiryanov98}.
For Al-doped Fe$_3$Ga$_4$, Duijn et al.~\cite{Duijn00} described a similar \emph{H-T} phase diagram in a similar manner.
In such a picture, the phase SF is canted, with a non-zero magnetization, and, moreover, is characterized by a, possibly additional, wave vector different from the zero-field wave vector. This could possibly be verified by a neutron-diffraction experiment.    

In a similar way, the \emph{H-T} phase diagram for undoped \CeFe\ can be evaluated. Here, at temperatures above \TT, there exists, besides the zero-field phase AF1, two additional ordered phases, SF1 and SF3. If the introduction of an additional competing magnetization for Ta-doped \CeFe\ holds, both SF1 and SF3 are canted, and even more modes \emph{M}$_Q$ are relevant here.
Within the framework of the Moriya-Usami theory, a small amount of (Ta) dopant influences the competition of \emph{M}$_0$ and relevant \emph{M}$_Q$s and the number of relevant \emph{M}$_Q$ in the intermediate-temperature ordered state. 
As indicated in the introduction, the electronic density of states is probably strongly peaked near the Fermi-surface, and therefore the electronic and magnetic properties are likely strongly dependent on volume. Different intermediate-temperature behavior of Ta-doped \CeFe\ and undoped \CeFe\ may well be mostly related to differences in volume.

\section{Conclusions}

The magnetism of \CeFe, earlier thought to be abnormal, may be explained as being due to competing ferro- and antiferromagnetism, and is well described by a Moriya-Usami-type phenomenological theory, which in its unmodified form yields magnetic phase diagrams for systems with a competition between one antiferromagnetic and one ferromagnetic mode. 
Magnetic phase diagrams obtained for undoped \CeFe\ indicate that the magnetic order is determined by the competition between multiple antiferromagnetic modes and a ferromagnetic mode. This competition and the resulting magnetic order are sensitive to minor changes in the crystallographic unit cell. A minor amount of Ta preferentially substitutes for Fe on one of the 4 Fe crystallographic sites and enlarges the unit-cell volume. 
This Ta doping influences both the competition and the number of competing magnetic modes, and results in a ferromagnetic ground state.

Within the ordered state of undoped \CeFe, as detailed elsewhere in this volume\cite{Kreyssig06}, a doubling of the unit cell in the \emph{c}-direction takes place, which is due to a minor atomic displacement. Also this crystallographic modification has a distinct effect on both the competition and the number of competing magnetic modes. Here, the resulting ground state is antiferromagnetic, and it may thus be said that the crystallographic phase transition stabilizes the antiferromagnetism. The sensitivity of the magnetism to Ta doping and to a minor displacive change in crystal structure is likely related to a strong sensitivity of electronic properties to small changes in (local atomic) volume, which in turn is consistent with a strongly peaked density of states near the Fermi level, as reported for the related compound \YFe. 

In conclusion, the present work indicates that the same mechanisms that determine the magnetic properties of \RFe\ compounds, the presence of both ferro- and antiferromagnetic exchange interactions, and lattice anomalies, play a crucial role in determining the magnetism in \CeFe.

\acknowledgments

The authors are indebted to S. L. Bud'ko, R. W. McCallum, K. W. Dennis, J. Frederick, S. Jia, M. Angst, R. J. McQueeney, and A. I. Goldman. Ames Laboratory is operated for the U.S. Department of Energy by Iowa State University under Contract No. W-7405-ENG-82.  This work was supported by the Director for Energy Research, Office of Basic Energy Sciences.

\end{document}